\documentclass{llncs}
\usepackage{amsmath}
\usepackage{amssymb}
\usepackage{enumerate}
\newcommand{\compl}[1]{\ensuremath{\overline{#1}}}
\newcommand{\boxi}{\ensuremath{\mathrm{box}}}
\newcommand{\cubi}{\ensuremath{\mathrm{cub}}}
\newcommand{\bw}{\ensuremath{\mathrm{bw}}}

\newcommand {\bbox}{\rule{0.6em}{0.6em}}
\newcommand{\ignore}[1]{}

\newcommand{\abs}[1]{\ensuremath{\left|#1\right|}}

\newcommand{\mc}[1]{\ensuremath{\mathcal #1}}

\newcommand{\surl}[1]{{\small\url{#1}}}

\title  {On the Cubicity of AT-free graphs and Circular-arc graphs}

\author{L. Sunil Chandran \and Mathew C. Francis  
\thanks{Indian Institute of Science,
Dept. of Computer Science and Automation,
Bangalore--560 012, India.  email: \emph{mathew,sunil@csa.iisc.ernet.in}}
\and Naveen Sivadasan
\thanks{Advanced Technology Centre, TCS, Deccan Park, Madhapur,
Hyderabad--500 081, India. email: \emph{s.naveen@atc.tcs.com}}
\institute{}}

\begin{document}
\maketitle
\pagestyle{plain}
\pagenumbering{arabic}

\begin{abstract}
A unit cube in $k$ dimensions ($k$-cube) is defined as the 
the Cartesian product $R_1\times R_2\times\cdots\times R_k$ where $R_i$(for
$1\leq i\leq k$) is a closed interval of the form $[a_i,a_i+1]$ on the real
line.
A graph $G$ on $n$ nodes is said to be representable as the intersection of
$k$-cubes (cube representation in $k$ dimensions) if
each vertex of $G$ can be mapped to a $k$-cube such that two vertices are
adjacent in $G$ 
if and only if their corresponding $k$-cubes have a non-empty intersection.
The \emph{cubicity} of $G$ denoted as $\cubi(G)$ is the minimum $k$ for which
$G$ can be represented as the intersection of $k$-cubes. 

An interesting aspect about cubicity is that many problems known to be 
NP-complete for general graphs have polynomial time deterministic algorithms 
or have good approximation ratios in graphs of low cubicity. 
In most of these algorithms, computing a low dimensional cube representation
of the given graph is usually the first step.

We give an $O(bw\cdot n)$ algorithm to compute the cube representation of a
general graph $G$ in $bw+1$ dimensions given a bandwidth ordering of the
vertices of $G$, where $bw$ is the \emph{bandwidth} of $G$.
As a consequence, we get $O(\Delta)$ upper bounds on the cubicity of
many well-known graph classes such as
AT-free graphs, circular-arc graphs and co-comparability
graphs which have $O(\Delta)$ bandwidth. Thus we have:
\begin{enumerate}
\item $\cubi(G)\leq 3\Delta-1$, if $G$ is an AT-free graph.
\item $\cubi(G)\leq 2\Delta+1$, if $G$ is a circular-arc graph.
\item $\cubi(G)\leq 2\Delta$, if $G$ is a co-comparability graph.
\end{enumerate}
Also for these graph classes, there are constant factor approximation
algorithms for bandwidth computation
that generate orderings of vertices with $O(\Delta)$ width.
We can thus generate the cube representation of
such graphs in $O(\Delta)$ dimensions in polynomial time.
\vspace{0.2in}
\end{abstract}

\bibliographystyle{plain}

\section{Introduction}
Let $\mathcal{F}=\{S_x\subseteq U:x\in V\}$ be a family of subsets of a universe $U$, where $V$ is an index set. The intersection graph $\Omega(\mathcal{F})$ of $\mathcal{F}$ has $V$ as vertex set, and two distinct vertices $x$ and $y$ are adjacent if and only if $S_x\cap S_y\neq\emptyset$. Representations of graphs as the intersection graphs of various geometrical objects is a well studied topic in graph theory. Probably the most well studied class of intersection graphs are the \emph{interval graphs}, where each $S_x$ is a closed interval on the real line. A restricted form of interval graphs, that allow only intervals of unit length, are \emph{indifference graphs}.

A well known
concept in this area of graph theory is the \emph{cubicity}, which
was introduced by F. S. Roberts in 1969  \cite {Roberts}.
This concept generalizes the concept of indifference graphs.
A unit cube in $k$ dimensions ($k$-cube) is a Cartesian product 
$R_1 \times R_2 \times \cdots \times R_k$ where $R_i$ (for $1 \le i \le k$)
is a closed interval of the form $[a_i, a_i + 1]$ on the real line. 
Two $k$-cubes, $(x_1,x_2,\ldots,x_k)$ and $(y_1,y_2,\ldots,y_k)$ are said to have a non-
empty intersection if and only if the intervals $x_i$ and $y_i$ have a non-empty intersection
for $1\leq i\leq k$.
For a  graph $G$, its \emph{cubicity}  is the minimum dimension $k$, such that
$G$ is representable as the intersection graph of $k$-cubes.
We denote the cubicity of a graph $G$ by $\cubi(G)$. 
The graphs of cubicity at most $1$ are exactly the class of indifference graphs.

If we require that each vertex correspond to a $k$-dimensional axis-parallel
box $R_1 \times R_2 \times \cdots \times R_k$ where $R_i$ (for $1 \le i \le k$)
is a closed interval of the form $[a_i, b_i]$ on the real line, then
the minimum dimension required to represent $G$ is called its $boxicity$
denoted as $\boxi(G)$. Clearly $\boxi(G) \le \cubi(G)$ for any graph $G$
because cubicity is a stricter notion than boxicity.

It has been shown that deciding whether the cubicity of a given graph is at
least 3 is NP-hard \cite{Yan1}.

In many algorithmic problems related to graphs, the availability of
certain convenient representations turn out to be extremely useful.
Probably, the most well-known and important examples are the tree decompositions
and path decompositions.
Many NP-hard problems are known to be
polynomial time solvable given a tree(path) decomposition  of the input
graph that has bounded width. Similarly, the representation of
graphs as intersections of ``disks" or ``spheres" lies at the core
of solving problems related to frequency assignments in radio networks,
computing molecular conformations etc. For the  maximum independent set problem
which is hard to approximate  within a factor of $n^{(1/2) - \epsilon}$
for general graphs, a PTAS is known for disk graphs given the disk representation  
\cite{Erl01,Chan01} and an FPTAS is known for unit disk graphs \cite{leeuwen05}. 
In a similar way,
the availability of cube or box representation in low dimension
make some well known NP hard problems like the max-clique problem, 
polynomial time solvable since there are only $O((2n)^k)$ maximal cliques
if the boxicity or cubicity is at most $k$.
Though the  complexity of finding
the maximum independent set is hard to approximate within a factor $n^{(1/2) - \epsilon}$
for general graphs, it is approximable to a $\log n$ factor for boxicity $2$
graphs (the problem is NP-hard  even for boxicity $2$ graphs) given a 
box or cube representation \cite{Agarwal98,Berman2001}.

It is easy to see that the problem of representing graphs using $k$-cubes can be equivalently formulated as
the following geometric embedding problem. Given an undirected unweighted graph
$G=(V,E)$ and a threshold $t$, find an embedding $f : V \rightarrow \mathbb{R}^{k}$
of  the vertices of $G$ into a $k$-dimensional space
(for the minimum possible $k$) such that for any two vertices $u$ and $v$ of $G$,
$||f(u) - f(v)||_{\infty} \le t$ if and only if $u$ and $v$ are adjacent.
The norm $||~~||_{\infty}$ is the $L_{\infty}$ norm. Clearly,
a $k$-cube representation of $G$ yields the required embedding of $G$ in the $k$-dimensional space.
The minimum dimension required to embed $G$ as above under the $L_2$ norm
is called the \emph{sphericity} of $G$.
Refer \cite{quint2} for applications where 
such an embedding under $L_{\infty}$ norm is argued to be more appropriate than embedding under $L_2$ norm.
The connection between cubicity and sphericity of graphs were studied in \cite{Fishburn,Maehara}.

As far as we know, the only known upper bound for the cubicity of general 
graphs (existential or constructive)
is by Roberts \cite{Roberts}, who showed that $\cubi(G) \le 2n/3$
for any graph $G$ on $n$ vertices. 
The cube representation of special class of graphs like hypercubes and complete multipartite graphs 
were investigated in \cite{Roberts,Maehara,Quint}.

~~~~~~~~~~~~~~~~~~~~~~~~~~~~~~~~~~~~~~~~~~~~~~~~~~~~~~~~~~~~~~~~~~~~~~

\noindent
\emph{Linear Ordering and Bandwidth.} Given an undirected graph $G=(V,E)$
on $n$ vertices,  a \emph{linear ordering} of $G$ is a bijection
$f : V \rightarrow \{1,\ldots,n\}$.
The \emph{width} of the linear ordering $f$ is defined as $\max_{(u,v) \in E} |f(u) - f(v)|$.
The \emph{bandwidth minimization problem} is to compute $f$ with minimum possible width.
The \emph{bandwidth} of $G$ denoted as $\bw(G)$ is the minimum possible width achieved by any linear
ordering of $G$. A \emph{bandwidth ordering} of $G$ is a linear ordering of $G$
with width $\bw(G)$.
Our algorithm  to compute the cube representation of a graph $G$ takes as input 
a linear ordering of $G$. The smaller the width of this ordering, the lesser
the number of dimensions of the cube representation of $G$ computed by our algorithm.
It is NP-hard to approximate the bandwidth of $G$ within a ratio better than
$k$ for every $k \in \mathbb{N}$ \cite{Ung98}. 
Feige \cite{Feige98} gives a $O(\log^3(n)\sqrt{\log n \log\log n})$ approximation algorithm
to compute the bandwidth (and also the corresponding linear
ordering) of general graphs.  For bandwidth computation, several algorithms
 with good heuristics are known that perform very well in practice \cite{Turner}.

\subsection{Our results}
We summarize below the results of this paper.
\begin{enumerate}
\item For any graph $G$, $\cubi(G)\leq \bw(G)+1$
\item For an AT-free graph $G$ with maximum degree $\Delta$,
$\cubi(G)\leq 3\Delta-1$
\item For a circular-arc graph $G$ with maximum degree $\Delta$,
$\cubi(G)\leq 2\Delta+1$
\item For a co-comparability graph $G$ with maximum degree $\Delta$,
$\cubi(G)\leq 2\Delta$
\end{enumerate}

\subsection{Definitions and Notations}
All the graphs that we consider will be simple, finite and undirected.
For a graph $G$, we denote the vertex set of $G$ by $V(G)$ and the
edge set of $G$ by $E(G)$. For a vertex $u \in V$, let $d(u)$ denote its
degree (the number of outer neighbors of $u$).
The maximum degree of $G$ is denoted
by $\Delta(G)$ or simply $\Delta$ when the graph under consideration is clear.
For a vertex $u\in V(G)$, we denote the set of neighbours of $u$ by $N_G(u)$.
By definition, $N_G(u)=\{v\in V(G)~|~(u,v)\in E(G)\}$. Again, for ease of
notation, we use $N(u)$ instead of $N_G(u)$ when there is no scope for
ambiguity. Let $G'$ be a graph such that $V(G')=V(G)$. Then $G'$ is a 
\emph{supergraph} of $G$ if $E(G)\subseteq E(G')$. We define the 
\emph{intersection} of two graphs as follows.
If $G_1$ and $G_2$ are two graphs such that $V(G_1)=V(G_2)$, then the
intersection of $G_1$ and $G_2$ denoted as $G=G_1\cap G_2$ is the graph with
$V(G)=V(G_1)=V(G_2)$ and $E(G)=E(G_1)\cap E(G_2)$.

An indifference graph is an interval graph which has an interval
representation that maps the vertices to unit length intervals on the real
line such that two vertices are adjacent in the graph if and only if the 
intervals mapped to them overlap.

\begin{definition}[Unit interval representation]
Given an indifference graph $I(V,E)$, the unit interval representation
is a mapping $f: V \rightarrow \mathbb{R}$ such that for any two vertices
$u, v$, $\abs{f(u) - f(v)} \le 1$ if and only if $(u, v)\in E$.
\end{definition}

Note that this is equivalent to mapping each vertex of $I$ to the unit
interval $[f(u),f(u)+1]$ so that two vertices are adjacent in $I$ if and only
if the unit intervals mapped to them overlap. Now, consider the mapping
$g:V\rightarrow \mathbb{R}$ given by $g(u)=x f(u)$ where $x\in \mathbb{R}$.
It can be easily seen that
for any two vertices $u, v$, $\abs{g(u)-g(v)}\le x$ if and only if $(u,v)$ is
an edge in $I$. $g$ thus corresponds to an interval representation of $I$
using intervals of length $x$. We call such a mapping $g$
a \emph{unit interval representation of $I$ with interval length $x$}.
\begin{definition}[Indifference graph representation] \label{def_indiff}
The indifference graphs $I_1, \ldots, I_k$ constitute an indifference
graph representation of a graph $G$ if $G = I_1 \cap \cdots \cap I_k$.
\end{definition}
\begin{theorem}[Roberts\cite{Roberts}]\label{indiffrepresentation}
A graph $G$ has $\cubi(G)\leq k$ if and only if it has an indifference graph
representation with $k$ indifference graphs.
\end{theorem}

\section{Cubicity and bandwidth}
\subsection{The construction}
We show that given a linear ordering of the vertices of $G$ with width
$b$, we can construct an indifference graph representation of $G$ using
$b+1$ indifference graphs.

\begin{theorem}\label{cubband1}
If $G$ is any graph with bandwidth $b$, then $\cubi(G)\leq b+1$.
\end{theorem}
\begin{proof}
Let $n$ denote $|V(G)|$ and let $\mathcal{A}=u_1,u_2,\ldots,u_n$ be a linear
ordering of the vertices of $G$ with width $b$. i.e., if $(u_i,u_j)\in E(G)$, 
then $\abs{i-j}\leq b$.

We construct $b+1$ indifference graphs $I_0,I_1,\ldots,I_b$, such that
$G=I_0\cap I_1\cap\cdots\cap I_b$. Let $f_i$ denote the unit interval
representation of $I_i$.\vspace{0.05in}

\noindent\textbf{Construction of $I_0$:}

Since $I_0$ has to be a supergraph of $G$, we have to make sure that every
edge in $E(G)$ has to be present in $E(I_0)$. $b$ being the bandwidth of
the linear ordering \mc{A} of vertices taken, a vertex $u_j$ is not adjacent
in $G$ to any vertex $u_k$ when $\abs{j-k}>b$. Now, we define $f_0$
in such a way that $E(I_0)=\{(u_j,u_k)~|~\abs{j-k}<b\}\cup\{(u_j,u_k) ~|~ 
\abs{j-k}=b\mbox{ and }(u_j,u_k)\in E(G)\}$.
The definition of $f_0$ can be explained as the following procedure.
We first assign the interval $[j,j+b]$ to vertex $u_j$, for all $j$.
This makes sure
that $u_j$ is not adjacent to any vertex $u_k$, if $k>j+b$. Now, each
vertex is adjacent in $I_0$ to exactly the $b$ vertices preceding and following
it in \mc{A}. Now, for each vertex $u_j$ where $j>b$, we shift $f_0(u_j)$, 
the unit interval for $u_j$, slightly to the right (by $\epsilon$) 
if $u_j$ is not adjacent to $u_{j-b}$
in $G$ so that $f_0(u_j)$ becomes disjoint from $f_0(u_{j-b})$. Along with
$f_0(u_j)$, all the intervals that start after $f_0(u_j)$ are also shifted right
by $\epsilon$. This procedure is done for vertices $u_{b+1},\ldots,u_n$ in
that order. Our choice of
a small value for $\epsilon$ ensures that $I_0$ is still a supergraph of $G$.

$f_0$ is a unit interval representation for $I_0$ with interval length $b$
defined as follows. Let $\epsilon=1/n^2$.
\begin{eqnarray*}
f_0(u_j)&=&j, \mbox{ for } j\leq b\\
f_0(u_j)&=&f(u_{j-b})+b, \mbox{ for } j>b\mbox{ and } (u_{j-b},u_j)\in E(G)\\
f_0(u_j)&=&f(u_{j-b})+b+\epsilon, \mbox{ for } j>b\mbox{ and } 
(u_{j-b},u_j)\not\in E(G)
\end{eqnarray*}

\noindent\textbf{Construction of $I_i$, for $1\leq i\leq b$ :}

We split the sequence of vertices \mc{A} into blocks $B_0^i,\ldots,B_{p-1}^i$
of vertices of size $b$ starting from the vertex $i$ where the last block
$B_{p-1}^i$ may have less than $b$ vertices. Formally,
$B_t^i=\{u_{i+bt},\ldots,u_{i+b(t+1)-1}\}$, for $1\leq t< p-1$,
and $B_{p-1}^i=\{u_{i+b(p-1)},\ldots,u_n\}$. Let $s_t^i$ denote the vertex
$u_{i+bt}$, or the first vertex (in the ordering \mc{A}) in block $B_t^i$.
We now define $f_i$, the unit interval representation for $I_i$ with interval
length 2, as follows:\\
$$f_i(u_j)=2\mbox{, if }j<i$$\\
Let $u$ be a vertex in $V(G)-\{u_j ~|~ j<i\}$ and let $u\in B_t^i$.
\begin{eqnarray*}
f_i(u)&=&t\mbox{, if }u=s_t^i\\
&=&t+2\mbox{, if }(u,s_t^i)\in E(G)\\
&=&t+3\mbox{, if }(u,s_t^i)\not\in E(G)
\end{eqnarray*}

\begin{claim}
$I_0$ is an indifference supergraph of $G$.
\end{claim}
\begin{proof}
First we observe that for any vertex $u_j$, $j \le f_0(u_j) \le j + 1/n$.
This is because  $f_0(u_j) \le f_0(u_{j-b}) + b + \epsilon$ where 
$\epsilon = 1/n^2$.
Now, consider an edge $(u_j, u_k)$ of $G$ where $j < k$.
Since the width of the input linear ordering $\mathcal{A}$ is $b$, we 
have
$k - j \le b$.
Now we consider the following two cases.
If $k-j \le b-1$ then $f(u_k) - f(u_j) \le k + 1/n - j \le b-1 + 1/n < b$.
Since each interval in $I_0$ has length $b$, it follows that $(u_j, u_k) 
\in E(I_0)$.
If $k-j =b$ then from the definition of $f_0$, it follows that
$f_0(u_k) = f_0(u_{k-b}) + b = f_0(u_j) + b$. Thus $f_0(u_k) - f_0(u_j) 
\le b$
implying that $(u_j, u_k) \in E(I_{0})$.
\end{proof}

\begin{claim}
$I_i$ for $1 \le i \le b$ is an indifference supergraph of $G$.
\end{claim}
\begin{proof}

Consider the indifference graph $I_i$. Let $(u_j,u_k)$ be any edge
in $E(G)$. We assume without loss of generality that $j<k$. If $j<i$,
then $k<i+b$ and therefore, $u_k\in B_0^i$. In this case, $f_i(u_j)=2$ and
$0\leq f_i(u_k)\leq 3$ and so we have $\abs{f_i(u_j)-f_i(u_k)}\leq 2$.
Now, consider the case when $j\geq i$. Let $u_j\in
B_t^i$. Since $\abs{j-k}\leq b$, we have $u_{k}\in B_t^i \cup B_{t+1}^i$.
From the definition of $f_i$, it is clear that $\abs{f_i(u_j)-f_i(u_k)}\leq 2$
if $u_j\not=s_t^i$. Now, if $u_j=s_t^i$, then either $u_k\in B_t^i$,
in which case $f_i(u_k)=t+2$, or $u_k=s_{t+1}^i$, in which case $f_i(u_k)
=t+1$. But in both cases, $\abs{f_i(u_j)-f_i(u_k)}\leq 2$. Therefore,
we have $f_i(u_j)\cap f_i(u_k)\not=\emptyset$ which implies that
$(u_j,u_k)\in E(I_i)$.

\end{proof}
It remains to show that $G=I_0 \cap \cdots \cap I_b$. To do this, it 
suffices to show that for any $(u_j, u_k) \notin E(G)$,
there exists an $I_i$, $0 \le i \le b$ such that $(u_j, u_k) \notin 
E(I_i)$. Let $j < k$.
Case $k - j \ge b$. In this case, we claim that $(u_j, u_k) \notin 
E(I_0)$. This is because of the following.
If $k -j = b$ then
clearly $f_0(u_k) - f_0(u_j) = b + \epsilon$ and thus $(u_j, u_k) \notin 
E(I_0)$.
Now, if $k - j > b$ then $f_0(u_k) \ge f_0(u_{k - b}) + b > f_0(u_j) + b$ 
observing that
$f_0(u_1) < f_0(u_2) < \cdots < f_0(u_n)$. Thus $(u_j, u_k) \notin 
E(I_0)$.
Now the remaining case is $k - j < b$. Consider the graph $I_l$ where $l = 
j \mod b$. Let $t$ be such that $u_j\in B_t^l$. Therefore, $bt+l\leq j<
b(t+1)+l$. This implies that $bt+l=bt+ j\mod b=j$ and so we have $s_t^l=u_j$.
Therefore, $u_k\in B_t^l$ since $k<j+b=b(t+1)+l$.
Now, from the definition of $f_l$, we have $f_l(u_j)=t$ and $f_l(u_k)=t+3$.
Thus, $\abs{f_l(u_j)-f_l(u_k)}>2$ and hence $(u_j, u_k) \notin E(I_l)$ as 
required.

Thus $I_0,\ldots,I_b$ is a valid indifference graph representation of $G$
using $b+1$ indifference graphs which establishes that $\cubi(G)\leq b+1$.
\hfill\bbox
\end{proof}
\subsection{The algorithm}
Our algorithm to compute the cube representation of $G$ in $b+1$ dimensions
given a linear ordering of the vertices of $G$ with width $b$ constructs 
the indifference supergraphs of $G$, namely, $I_0,\ldots,
I_b$ using the constructive procedure used in the proof of Theorem
\ref{cubband1}. It is easy to verify that this algorithm runs in
$O(b\cdot n)$ time where $b$ is the width of the input linear arrangement
and $n$ is the number of vertices in $G$.
\section{Applying our results}
Theorem \ref{cubband1} can be used to derive upper bounds for the cubicity of
several special classes of graphs such as circular arc graphs,
co-comparability graphs and AT-free graphs.

\begin{corollary}
If $G$ is a circular-arc graph, $\cubi(G)\leq 2\Delta+1$, where $\Delta$ is
the maximum degree of $G$.
\end{corollary}
\begin{proof}
Let an arc on a circle corresponding to a vertex $u$ be denoted by $[h(u),
t(u)]$ where $h(u)$(called the \emph{head} of the arc) is the starting point
of the arc when the circle is traversed in the clockwise order and $t(u)$
(called the \emph{tail} of the arc) is the ending point of the arc when
traversed in the clockwise order. We assume without loss of generality that 
the end-points of all the arcs are distinct and that no arc covers the whole
circle. If any of these cases occur, the end-points of the arcs can be shifted
slightly so that our assumption holds true.

Choose a vertex $v_1\in V(G)$. Start from $h(v_1)$ and traverse the circle in
the clockwise order. We order the vertices of the graph (other than $v_1$) as
$v_2,\ldots,v_n$ in the order in which the heads of their corresponding arcs
are encountered during this traversal.
Now, we define an ordering $f:V(G)\rightarrow \{1,\ldots,n\}$ of the
vertices of $G$ as follows:\\
$f(v_j)=2j$, if $1\leq j\leq \lfloor n/2\rfloor$.\\
$f(v_j)=2(n-j)+1$, if $\lfloor n/2\rfloor<j\leq n$.\\
We now prove that the width of this ordering is at most $2\Delta$. 

We claim that if $h(v_j)$ and $h(v_k)$ are two consecutive heads encountered
during a clockwise traversal of the circle, $|f(v_j)-f(v_k)|\leq 2$. To see
this, we will consider the different cases that can occur:\\
Case : When $1\leq j<j+1=k\leq\lfloor n/2\rfloor$. Here, $f(v_j)=2j$ and
$f(v_k)=2(j+1)$. Therefore, $|f(v_j)-f(v_k)|=2$.\\
Case : When $\lfloor n/2 \rfloor<j<j+1=k\leq n$. In this case, $f(v_j)=
2(n-j)+1$ and $f(v_k)=2(n-(j+1))+1$, which means that $|f(v_j)-f(v_k)|=2$.\\
Case : When $j=\lfloor n/2\rfloor<j+1=k$,\\
~~Subcase : If $n$ is even. $f(v_j)=2j=n$ and $f(v_k)=2(n-(j+1))+1=2n-2j-1
=n-1$.\\
~~Subcase : If $n$ is odd, $f(v_j)=2j=n-1$ and $f(v_k)=2n-2j-1=n$.\\
In both these cases, $|f(v_j)-f(v_k)|=1$.\\
Case : When $j=n$ and $k=1$. We then have $f(v_j)=1$ and $f(v_k)=2$.
Therefore, $|f(v_j)-f(v_k)|=1$.

Now, consider any edge $(v_j,v_k)\in E(G)$. Assume without loss of generality
that $h(v_j)$ occurs first
when we traverse the circle in clockwise direction starting from $h(v_1)$.
Now, if we traverse the arc corresponding to
$v_j$ from $h(v_j)$ to $t(v_j)$, we will encounter at most $\Delta-1$ heads
$h(u_1),h(u_2),\ldots,h(u_{\Delta-1})$ before we reach $h(v_k)$ since $v_j$ can be connected to at most
$\Delta$ vertices in $G$. We already know that $|f(v_j)-f(u_1)|\leq 2$ and
$|f(u_i)-f(u_{i+1})|\leq 2$, for $1\leq i\leq \Delta-2$. Also,
$|f(u_{\Delta-1}-f(v_k)|\leq 2$. It follows that
$|f(v_j)-f(v_k)|\leq 2\Delta$.
Thus $f$ is an ordering of the vertices of $G$ with width at most $2\Delta$.
It follows from theorem \ref{cubband1} that $\cubi(G)\leq 2\Delta+1$.

\end{proof}

\begin{corollary}
If $G$ is a co-comparability graph, then $\cubi(G)\leq 2\Delta$, where $\Delta$
is the maximum degree of $G$.
\end{corollary}
\begin{proof}
Let $V$ denote $V(G)$ and let $|V|=n$.
Since $\compl{G}$ is a comparability graph, there exists a
partial order $\prec$ in $\compl{G}$ on the node set $V$ such that $(u,v)
\in E(\compl{G})$ if and only if $u\prec v$ or $v\prec u$. This partial order
gives a direction to the edges in $E(\compl{G})$. We can run a topological
sort on this partial order to produce a linear ordering of the vertices, say,
$f:V\rightarrow \{1,\ldots,n\}$. The topological sort ensures that if
$u\prec v$, then $f(u)<f(v)$. Now, let $(u,v)\in E(G)$ and let $w$ be a vertex
such that $f(u)<f(w)<f(v)$. We will show that $w$ is adjacent to either $u$
or $v$ in $G$. Suppose not. Then $(u,w),(w,v)\in E(\compl{G})$ and therefore
$u\prec w$ and $w\prec v$. Now, by transitivity of $\prec$, this implies that
$u\prec v$, which means that $(u,v)\in E(\compl{G})$ -- a contradiction.
Therefore, any vertex $w$ such that $f(u)<f(w)<f(v)$ in the ordering $f$ is
adjacent to either $u$ or $v$. Since the maximum degree of $G$ is $\Delta$,
there can be at most $2\Delta-2$ vertices between with $f(\cdot)$ value between
$f(u)$ and $f(v)$. Thus, the width of the ordering given by $f$ is at most
$2\Delta-1$ and by Theorem \ref{cubband1}, we have our bound on cubicity.
\end{proof}

A \emph{caterpillar} is a tree such that a path (called the \emph{spine}) is 
obtained by removing all its leaves. In the proof of Theorem 3.16 of
\cite{Klok6}, Kloks et al. show that every connected AT-free graph $G$ has a
spanning caterpillar subgraph $T$, such that adjacent nodes in $G$ are at a 
distance at most four in $T$. Moreover, for any edge $(u,v)\in E(G)$ such that
$u$ and $v$ are at distance exactly four in $T$, both $u$ and $v$ are leaves
of $T$. Let $p_1,\ldots,p_k$ be the nodes along the spine of $G$.

\begin{corollary}
If $G$ is an AT-free graph, $\cubi(G)\leq 3\Delta-1$, where $\Delta$ is the
maximum degree of $G$.
\end{corollary}
\begin{proof}
Let $L_i$ denote the set of leaves of $T$ adjacent to $p_i$. Clearly,
$|L_i|\leq \Delta$ and $L_i\cap L_j=\emptyset$ for $i\not=j$.
For any set $S$ of vertices, let $\langle S\rangle$ denote an arbitrary ordering of
the vertices in set $S$. Let $<u>$ denote ordering with just one vertex $u$
in it. If $\alpha=u_1,\ldots,u_s$ and $\beta=v_1,\ldots,
v_t$ are two orderings of vertices in $G$, then let $\alpha\diamond\beta$
denote the ordering $u_1,\ldots,u_s,v_1,\ldots,v_t$.
Let $\mc{A}=<L_1>\diamond <p_1>\diamond <L_2>\diamond <p_2>\diamond\cdots
\diamond <L_k>\diamond <p_k>$ be a linear ordering of the vertices of $G$.
One can use the property of $T$ stated before the theorem to easily show that
\mc{A} is a linear ordering of the vertices of $G$ with width at most
$3\Delta-2$. The corollary will then follow from Theorem \ref{cubband1}.
\end{proof}

\end{document}